# An anomalous magnetic phase transition at 10 K in $Nd_7Rh_3$

Kausik Sengupta and E.V. Sampathkumaran

*Tata Institute of Fundamental Research, Homi Bhabha Road, Colaba, Mumbai 400005, India*

The compound, $Nd_7Rh_3$, crystallizing in $Th_7Fe_3$-type hexagonal structure, has been shown recently by us to exhibit a signature of 'magnetic phase-coexistence phenomenon' below 10 K after a field cycling, uncharacteristic of stoichiometric intermetallic compounds, bearing a relevance to the trends in the field of 'electronic phase-separation'. In order to characterize this compound further, we have carried out dc magnetic susceptibility ($\chi$), electrical resistivity, magnetoresistance and heat-capacity measurements as a function temperature (T= 1.8 to 300 K). The results reveal that this compound exhibits another unusual finding at the 10K-transition in the sense that the plot of $\chi(T)$ shows a sharp increase in the 'field-cooled' cycle, whereas the 'zero-field-cooled' curve shows a downturn below the transition. In addition, the sign of magnetoresistance is negative and the magnitude is large over a wide temperature range in the vicinity of magnetic ordering temperature, with a sharp variation at 10 K. The results indicate that the transition below 10 K is first-order in its character.

# I. INTRODUCTION

Recently, we have reported [1] isothermal magnetization (M) and electrical resistivity ($\rho$) behavior as a function of externally applied magnetic field (H) for the binary intermetallic compound, $Nd_7Rh_3$ (Ref. 2), crystallizing in $Th_7Fe_3$-type hexagonal structure [3]. This compound exhibits two magnetic phase transitions, one at (T1= ) 32 K and the other at about (T2=) 10 K. On the basis of isothermal magnetization behavior, we have proposed [1] that both the transitions are essentially antiferromagnetic (AF) in character. The most intriguing property of this compound is that there is a magnetic-field-induced first-order antiferromagnetic-to-ferromagnetic transition around 10 kOe appearing below T2 and that there is a co-existence of antiferromagnetic and 'super-cooled' ferromagnetic (F) phases after the magnetic field is reduced zero, however, without spin-glass anomalies. The phase co-existence phenomenon for a stoichiometric intermetallic compound, mimicking the behavior of manganites, is unusual and bears significant relevance to the modern topic of 'electronic phase-separation' [4,5] as discussed in our earlier article [1]. We also found evidence for the 'memory' of high-field conductivity at low temperatures after the magnetic field is reduced to zero, which is taken as an evidence for percolative electrical conduction through less-resistive ferromagnetic-clusters dispersed in more-resistive antiferromagnetic-clusters. In this article, we present the results of dc magnetic susceptibility ($\chi$) and $\rho$ as a function of temperature (T) in the presence of various external fields as well as heat-capacity (C) measurements for this compound to bring out further interesting aspects of this compound.

## II. EXPERIMENTAL DETAILS

The polycrystalline sample employed in the present investigation is the same as that in our previous studies [1]. Temperature dependent (1.8-300 K) dc magnetization measurements in the presence of several fixed magnetic fields (100 Oe, 500 Oe, 1 kOe, and 5 kOe) were performed employing a commercial (Quantum Design) superconducting quantum interference device (SQUID) as well as by a vibrating sample magnetometer (VSM) (Oxford Instruments). The $\rho(T,H)$ behavior (1.8-300 K) was obtained with the help of a Physical Property Measurements System (PPMS) (Quantum Design). The same commercial set-up was used to measure heat-capacity by a relaxation method; in addition, the C(T) behavior was obtained by an adiabatic heat-pulse method employing a home-made set-up.

## III. RESULTS AND DISCUSSION
### A. Magnetic susceptibility

The results of dc $\chi$ measurements in the presence of various dc magnetic fields are shown in figure 1. ZFC-curves represent 'zero-field-cooled' data obtained while warming from 1.8 K in the presence of a desired field after cooling the specimen to 1.8 K in zero field, while field-cooled-warming (FCW) curves were obtained after cooling the sample in the presence of a field. The data were collected during cooling as well for a field of 100 Oe. One of the insets shows the plot of inverse $\chi$ obtained in a field of 5 kOe and this plot is linear above 40 K typical of paramagnets; the value of the effective moment is nearly the same (~3.72 $\mu_B$/Nd ion) as that of free $Nd^{3+}$ ion and the sign of paramagnetic Curie temperature (23 K) is positive indicating the existence of ferromagnetic correlations. There is a distinct peak in $\chi(T)$ in the vicinity of 30 K, indicative of the onset of long-range magnetic order. It should be noted that the peak appears at 32 K for H= 100 Oe, whereas, for H= 5 kOe, the peak appears at a slightly lower temperature (30 K). This implies a marginal suppression of magnetic transition temperature by the application of magnetic field, thereby establishing that the magnetic ordering near 32 K is of an antiferromagnetic type, despite a positive value of paramagnetic Curie temperature. ZFC-FCW curves do not show any bifurcation near 32 K and hence spin-glass freezing needs to be ruled out. At lower temperatures, we see an additional broad feature around 20 K for H= 100 Oe, which is gradually broadened as magnetic field is increased. There is a distinct bifurcation of ZFC-FCW curves around 20 K. It appears that, in addition to the 10K-magnetic transition as reported in the literature [2], there could be another magnetic transition around 20 K. The 20K-feature was not reported in the previous literature [2]. It may be recalled that there are three distinct crystallographically equivalent sites for Nd [3] and it is not clear whether the three different types of Nd ions order magnetically at different temperatures or whether there is a collective ordering at 32 K with spin-reorientation effects with decreasing temperature. The most noteworthy findings in the plots shown in figure 1 are: (i) The upturn in $\chi(T)$ at the onset of 10K-transition is quite sharp for a low field of 100 Oe, as though the transition is first order; however, there is a gradual broadening of the transition with increasing H; (ii) FCW curve for H= 100 Oe does not track ZFC curve and there is a dramatically sharp rise below 10 K, followed by saturation at lower temperatures; *to our knowledge, such a behavior of ZFC-FCW $\chi(T)$ curves was never witnessed in the literature;* as the magnetic field is increased, however, these two curves tend to track each other, eventually merging above 7 K for H= 5 kOe. Considering that the upturn below 10 K for FCW curve for H= 100 Oe is sharp, one is tempted to believe that the magnetic transition is first-order in its character in the presence of low fields. We have also obtained $\chi(T)$ curves while cooling in the presence of a magnetic field of 100 Oe and we found a very weak hysteresis (see the top inset in figure 1), as though this transition is first-order at low fields; we would however like to view this hysteretic effect with some degree of caution, as it was rather difficult to control the temperature to the desired accuracy while cooling the sample. The origin of the broadening of the transition at higher fields is at present puzzling to us.

**B. Heat capacity**

In figure 2, we show the results of heat capacity measurements to look for another evidence for magnetic transitions. The C(T) plot obtained with PPMS shows a well-defined anomaly at 32 K, whereas there is no feature at 10 K. In this connection, it is to be remembered [6] that the relaxation method employed in PPMS to measure heat capacity suppresses sharp features due to first-order transitions. In such cases, adiabatic heat-pulse method should be preferred to measure heat capacity. In fact, we have performed heat capacity measurements by a home-made set-up by semi-adiabatic heat-pulse method; the results are shown in the inset of figure 2, which reveal the existence of a well-defined anomaly at 10 K. This comparison of the heat capacity data obtained by these two methods appear to favor first-order nature of this 10K-transition. Finally, there is no prominent peak, but a weak one, around 20 K in C(T). Considering this, it is possible that the 20K-transition results in the low entropy change, though possible role of magnetic impurity contribution can not be ruled out.

**C. Electrical and magnetoresistance**

The results of temperature dependent electrical resistivity studies are shown in figure 3. The values of $\rho$ are typically much less than 1 m$\Omega$cm and d$\rho$/dT is positive down to 60 K in the entire temperature range of investigation. There is a drop at 32 and 10 K (for the data taken in zero field). All these findings are consistent with the previous report [2]. It should be noted that the drop at 10 K in our case is quite sharp as though the transition is first-order, whereas, in Ref.

2, the transition is found to be broad. We have also attempted to probe the influence of magnetic fields on the magnetic transitions and on the ρ(T) behavior, the results of which are shown (while warming) in figure 3 for the ZFC condition of the specimen. It is apparent from the figure 3 that, while the transition around 32 K gets broadened by the application of H, say 50 kOe, the feature at 10 K is washed out completely; this is due to the stabilization of ferromagnetism at high fields. In addition, the values of electrical resistivity are suppressed dramatically in the magnetically ordered state, resulting in negative magnetoresistance, defined as MR= [ρ(H)-ρ(0)]/ρ(0) (see figure 3, bottom). It is interesting to see that the magnitude of MR is as large as about 60% below 10 K, say for H= 50 kOe, arising from percolative conduction through ferromagnetic clusters as elaborated in Ref. 1; above 10 K also, values of MR are large arising from the field-induced ferromagnetism [1]. The magnitude of magnetoresistance peaks at the two (10 and 32 K) magnetic transitions. In fact, the suppression of ρ begins at a temperature which is nearly thrice of magnetic ordering temperature as in the case of heavy rare-earths of this series implications of which have been discussed at length in Refs. 7-9. Finally, there is no prominent peak close to 20 K neither in ρ(T) nor in MR(T), but the change of slopes in the curves (noticeable in MR(T) ) could possibly indicate a transition.

## IV. SUMMARY

The temperature dependence of magnetic susceptibility, magnetoresistance and heat-capacity of the compound, $Nd_7Rh_3$, which was recently reported to show novel 'phase co-existence' phenomenon following a field-cycling at low temperatures, is presented in this article. The results establish that there are at least two magnetic transitions, one at 32 K and the other at 10 K. The most interesting observation made here is that the transition at 10 K in low magnetic fields is found to be very sharp with an unusual bifurcation behavior of ZFC-FCW dc χ curves. It appears that the transition is first order. It is not clear whether there are any structural changes associated with this transition. Therefore, careful neutron and x-ray diffraction studies as a function of temperature and magnetic field will be quite rewarding.

**Acknowledgements**
The authors would like to thank Kartik K Iyer for his help in the measurements.

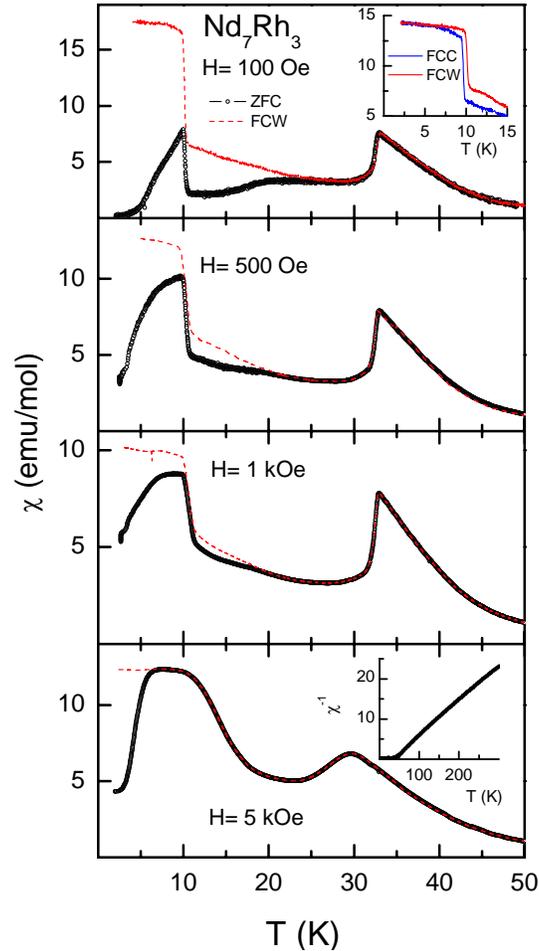

Fig. 1: (color online) The dc magnetic susceptibility ($\chi$) as a function of temperature (T) for $Nd_7Rh_3$ below 50 K, measured in the presence of various magnetic fields, for zero-field-cooled (ZFC, from 50 K)) and field-cooled (FCW) conditions of the specimens. The data shown are collected while warming. Top inset shows the profiles of FCC and FCW (field-cooled-warming) curves taken with H= 100 Oe to show hysteresis. Bottom inset shows inverse $\chi$ as a function of T below 300 K, obtained in a dc field of 5 kOe.

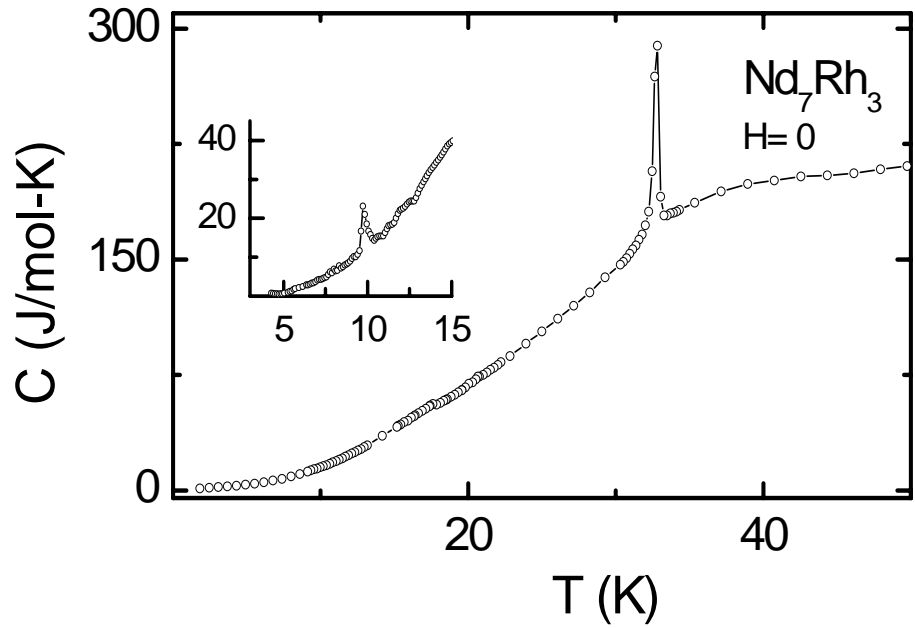

Fig. 2:
Heat-capacity as a function of temperature taken in the absence and in the presence (10 kOe) of a dc magnetic field, obtained by relaxation method with PPMS, for Nd$_7$Rh$_3$. The inset shows the data obtained by a semi-adiabatic heat-pulse method by a home-made calorimeter to highlight the existence of a transition at 10 K.

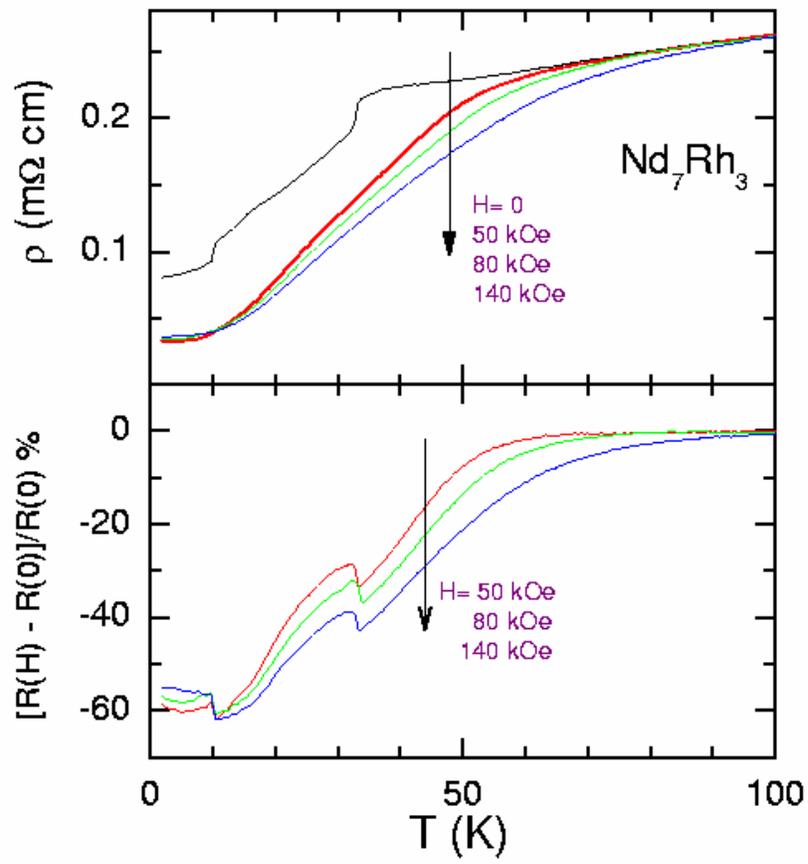

Fig. 3:
(color online) Electrical resistivity as a function of temperature below 100 K in the presence of various dc magnetic fields, for $Nd_7Rh_3$ (top). The magnetoresistance obtained from this data are plotted in the bottom figure.